\documentclass[amsmath,trackchanges, twocolumn]{aastex702}
\usepackage{graphicx} 
\usepackage{natbib}
\usepackage{amsmath}

\usepackage{gensymb}
\RequirePackage{mathtools}
\RequirePackage{gensymb}

\begin{document}

\title{Gone Without a Bang: A Search for Failed Core-Collapse Supernovae in ZTF and Predictions for Rubin and Roman}

\author[0009-0001-8820-7596]{Shion Murakawa}
\affiliation{MIT-Kavli Institute for Astrophysics and Space Research, 77 Massachusetts Ave., Cambridge, MA 02139, USA}
\email{shionmu@mit.edu}

\author[0000-0002-2218-2306]{Paul Draghis}
\affiliation{MIT-Kavli Institute for Astrophysics and Space Research, 77 Massachusetts Ave., Cambridge, MA 02139, USA}
\email{pdraghis@space.mit.edu}

\author[0000-0002-7226-836X]{Kevin Burdge}
\affiliation{MIT-Kavli Institute for Astrophysics and Space Research, 77 Massachusetts Ave., Cambridge, MA 02139, USA}
\email{kburdge@mit.edu}

\author[0000-0003-3115-2456]{Ore Gottlieb}
\affiliation{MIT-Kavli Institute for Astrophysics and Space Research, 77 Massachusetts Ave., Cambridge, MA 02139, USA}
\email{ogot@mit.edu}

\author[0000-0002-5812-3236]{Aaron Householder}
\affiliation{Department of Earth, Atmospheric and Planetary Sciences, Massachusetts Institute of Technology, Cambridge, MA 02139, USA}
\affiliation{MIT-Kavli Institute for Astrophysics and Space Research, 77 Massachusetts Ave., Cambridge, MA 02139, USA}
\email{aaron593@mit.edu}

\author[0000-0003-4780-4105]{Emma Chickles}
\affiliation{MIT-Kavli Institute for Astrophysics and Space Research, 77 Massachusetts Ave., Cambridge, MA 02139, USA}
\email{echickle@mit.edu}


\begin{abstract}
The core-collapse of massive stars (${\geq}10\ \rm M_\odot$) usually takes the form of a supernova, but in some cases may lead to a direct collapse, leading to a failed core-collapse supernova (FccSN). Wide-field time-domain surveys, such as the Zwicky Transient Facility (ZTF), are a useful tool to find such ``disappearing" stars. We use a piecewise linear function model to search the entire ZTF optical photometric catalog for FccSNe, yielding 125 ``faders" that dip significantly with a flat top and bottom flux. Although no candidates were definitively found, some of the sources exhibit promising behavior that may indicate a direct collapse event. Assuming a null result from the search, we calculate at 99\% credibility that the maximum statistical rate of FccSNe in the Milky Way is 147 per century. To contextualize these findings, we also simulate the formation of stellar populations in Milky Way-like galaxies to estimate the expected number of FccSNe observable by ZTF. This simulation predicts ZTF to observe $0.234\pm0.006$ FccSNe per century, which aligns with the lack of definitive observations in ZTF. However, we predict that similar monitoring using the Legacy Survey of Space and Time (LSST) and the Nancy Grace Roman Space Telescope's Wide Field Instrument (WFI) could enable population-level studies of nearby FccSNe.
\end{abstract}

\section{Introduction}

After reaching an iron core through nuclear fusion, massive stars (${>}10\ \rm M_\odot$) undergo a form of core collapse. The most well-studied form of core collapse is supernovae (ccSNe), most of which leave behind neutron stars (NSs) in their remnants \citep{pavlov_central_2004, 2011MNRAS.412L.108S, Heinz_2013}. Based on early analytical models, their progenitors are usually ${\sim}10$--$20\ \rm M_\odot$ stars \citep{Fryer_1999, Heger_2003}, with the lower bound corresponding to the lowest mass at which core-collapse will occur. Along the upper edge and above this mass range, core-collapse can result in a black hole (BH), although the outcome is also dependent on core mass, metallicity, rotation, and the progenitor's overall structure, and these numbers are quite model-dependent. Simulations in isolated systems (i.e. not in binaries) suggest that massive stars can undergo a quiet event to form BHs \citep{2011ApJ...730...70O}. These events are either 1: ``weak" supernova fallback accretion, where the explosion stalls and some of the ejecta fall back onto the proto-neutron star (PNS) remnant, exceeding the maximum mass supported by the PNS, or 2: a ``failed" supernova where the stalled explosion never restarts, causing complete gravitational collapse into a black hole \citep{2008ApJ...684.1336K}, also known as ``direct collapse". Models of isolated massive stars predict only ${\sim}10\%$ of successful ccSNe come from massive stars ${>}20\ \rm M_{\odot}$ \citep{2016ApJ...821...38S}, with stars $16.5{\sim}25\ \rm M_{\odot}$ being the primary source of failed core-collapse supernovae (FccSNe), corresponding to 20\% of all core collapses \citep{2014ApJ...785...28K}, although there are ``islands of explodability" within this mass range. 

Observationally, however, the line between direct collapse, supernovae, and compact object formation is unclear. ccSNe have missing progenitors in the mass range where either neutron stars or black holes can form. The Red Supergiant Problem is the observational absence of red supergiant supernova progenitors between 16.5 and 30 $\rm M_\odot$ \citep{2009ARA&A..47...63S}, and most missing progenitors come from the ${>}18\ \rm M_{\odot}$ threshold \citep{2015PASA...32...16S}. Furthermore, observations of supernova remnants are inconclusive of the core-collapse mechanism for more massive stars. Although most observed compact objects within supernova remnants are neutron stars \citep{pavlov_central_2004, 2011MNRAS.412L.108S, Heinz_2013}, some black holes candidates have also been found (e.g., \citealt{2023ApJ...947...38B}). 

Isolated black hole remnants of core-collapse are challenging to observe; most observationally confirmed black holes are observed in binary systems (e.g. X-ray binaries, gravitational waves from mergers; \citealt{Sahu_2022}). Although black holes in binaries are important observationally, they exacerbate the uncertainty in the core collapse mechanisms of massive stars \citep{2017A&A...601A..29Z, 2024ApJ...964...23W}. On the other hand, the actual collapse event can be easier to observe. There exist multiple supernovae surveys that cover a wide range of core-collapse events---e.g. The All-Sky Automated Survey for Supernovae (ASAS-SN) \citep{2017PASP..129j4502K} and the Dark Energy Survey supernova program (DES-SN) \citep{vincenzi2024darkenergysurveysupernova}. Direct collapse can be more difficult to search for than supernovae due to their inverse signal; however, some searches have yielded candidates for FccSN. \citet{2008ApJ...684.1336K} outlines a plan to monitor $10^6$ supergiants, and \citet{2017MNRAS.468.4968A} finds the first promising candidate N6946-BH1 with the Large Binocular Telescope, with its ``disappearance" also being observed by the Hubble Space Telescope (HST). Although a James Webb Space Telescope (JWST) follow-up by \citet{2024ApJ...964..171B} and \citet{2024ApJ...962..145K} found a faint remnant of the source in the mid-infrared (MIR) which appears similar to dusty stars, the significant drop in luminosity (${\sim}20\%$ of progenitor brightness) suggests otherwise. Another candidate M31-2014-DS1 was observed to disappear in the MIR by NEOWISE \citep{2026Sci...391..689D}. This source is believed to be a $12{\sim}13\ \rm M_\odot$ progenitor collapsing into a $5\ \rm M_\odot$ black hole (predicted through fallback models), with its disappearance confirmed through JWST follow-ups \citep{2026ApJ...999L..25D}. This source is more promising as a candidate since its proximity makes it much easier to study.

In this work, we set out to use data from the Zwicky Transient Facility (ZTF), a wide-field ($47\deg^2$) optical time-domain survey with a limiting g' magnitude of 20.8 and r' of 20.6 \citep{2019PASP..131a8002B, 2019PASP..131a8003M}, to search for FccSN. We use a set of filters and a logistic model to find evidence of ``disappearing" sources in the ZTF archive. Due to the angular resolution ($2"$) and limiting magnitude, this search will mostly be within the Milky Way. In accordance with this, we simulate theoretical expectation rates of FccSN within the Milky Way. Finally, we numerically set upper and lower bounds on FccSN rates with future observatories beyond our home galaxy.

\section{Fader Search on ZTF}

\subsection{Search Criteria}

We perform a search through ZTF light curves for FccSN candidates using a linear piecewise model. We use a catalog of forced photometry done on $1.5\times10^9$ Pan-STARRS \citep{2016arXiv161205560C} sources in the footprints on all g' and r' band ZTF images. First, a preliminary screening is done to filter out clear non-candidates since the fitting algorithm is both time-consuming and computationally expensive. We screen for sources that have their tenth percentile flux value less than the difference between their reference flux and 3 times their median error:
\begin{equation}
    \label{eq:lims}
    F_{\mathrm{10th}} < F_{\mathrm{ref}} - 3\bar{\sigma}
\end{equation}
This screen extracts for sources that have a significant drop in their reference value that could not be explained by noise. The search uses a linear piecewise function fitted to each ZTF light curve:
\begin{equation}
    \label{eq:16}
    F=\begin{cases}
        F_0 & t < t_0\\
        -\frac{\Delta F}{\Delta t}(t-t_0) +F_0 & t_0 \leq t < t_0+\Delta t\\
        F_0-\Delta F & t \geq t_0 + \Delta t
    \end{cases}
\end{equation}
where $F$ and $t$ are flux and time respectively, $F_0$ is the starting flux, $\Delta F$ is the Flux drop, $t_0$ is the start time of the drop, and $\Delta t$ is the duration of the drop. We limit $\Delta F$ and $\Delta t$ to be strictly positive to limit the model to only drop in flux over time. This piecewise model has a flat beginning and end, which is what we expect from a FccSN event. The source is further examined upon meeting these conditions:
\begin{enumerate}
    \item It has a $r^2$ value of ${>}0.75$.
    \item There are more than 5 epochs on either side of $t_0+\frac{\Delta t}{2}$.
    \item There are more than 10 epochs within $0.5\sigma$ of both the upper ($F_0$) and lower ($F_0-\Delta F$) flux bounds of the fitted logistic curve.
    \item The flux drop $\Delta F$ is ${>}5\sigma$ of fluxes $t<t_0$.
    \item The flux drop $\Delta F$ is over 25\% of the upper flux $F_0+L$.
\end{enumerate}

\begin{figure}[h!]
    \centering
    \includegraphics[width=\linewidth]{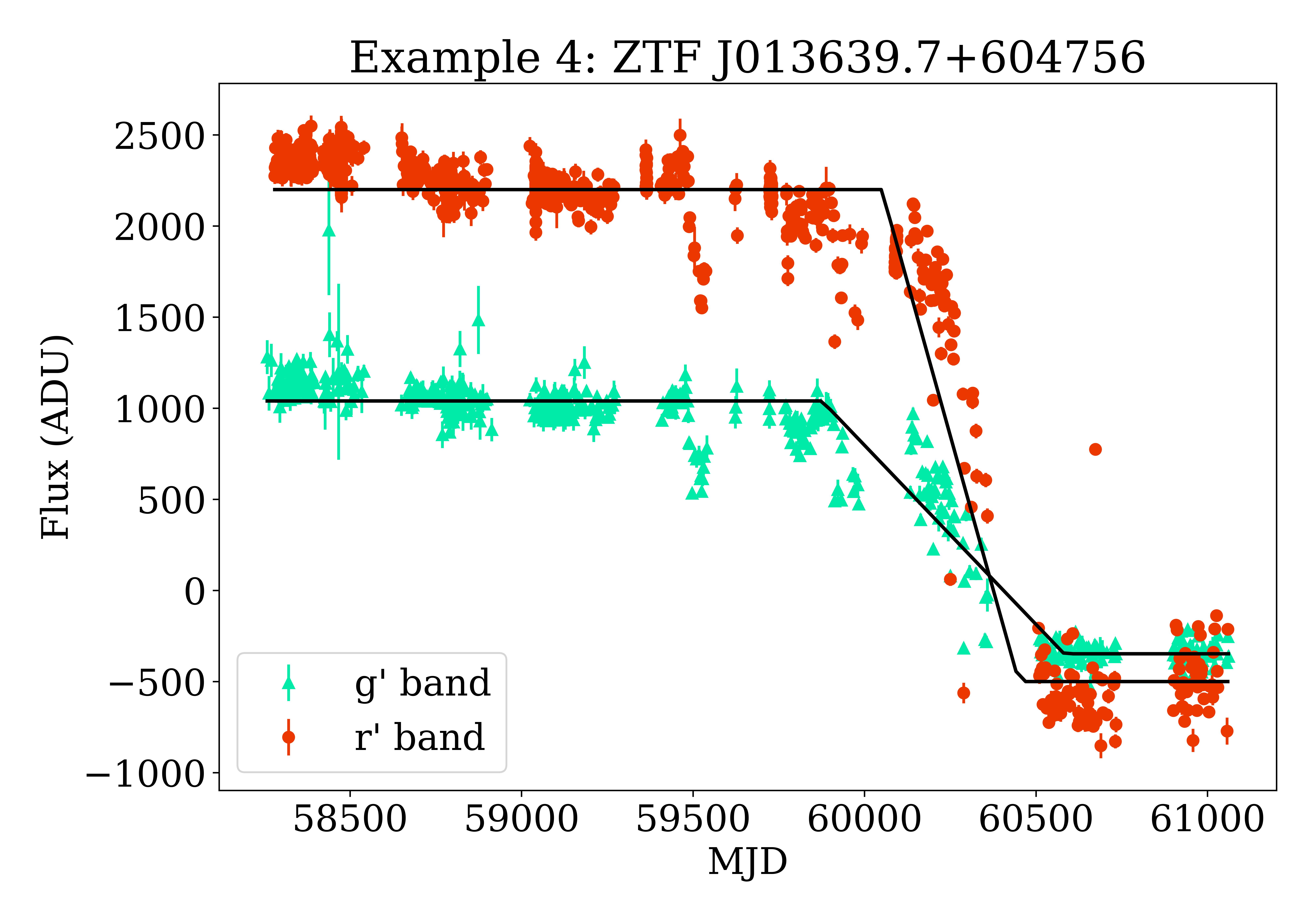}
    \caption{ZTF light curve of ZTF J013639.7+604756, a typical proto-candidate for a FccSN after the initial filters (model fit, HPM and QSO filter). The light curve is overlayed with the fitted linear model ($r^2_{g'}=0.93,\ r^2_{r'}=0.90$). Note that the negative flux is due to incorrect reference flux on file; fluxes are corrected after initial filters.}
    \label{fig:logmodel}
\end{figure}

If a source meets these conditions, we further filter it to rule out easily confirmable false positives. We cross-match the source with Gaia DR3 \citep{2023A&A...674A...1G} and discard sources with high proper motion (HPM) or active galaxy/quasi-stellar object (QSOs). HPM sources can mimic disappearing stars as they drift out of the fixed aperture; any source with a total proper motion ${>}30\ \mathrm{mas\ yr^{-1}}$ was discarded. Similarly, if an HPM source passes in front of the examined source (and starts layered on top during the first epoch), the flux will appear to decrease. However, these cases are more difficult to filter out due to the HPM source being different from the one being studied. Active galaxies can also have stochastic variability and can appear as if they are fading. A source passing all of these filters is shown in Figure \ref{fig:logmodel}; the source is relatively flat in flux until it drops over time, then begins to flatten again.

This initial search and cut yielded 769 sources, each of which was inspected by eye, mainly with the following criteria:
\begin{enumerate}
    \item The flux decrease appeared to be significant.
    \item It followed the flat-drop-flat shape, and not a linear drop for the entire light curve.
    \item The shape of the curve is not from some artifact of the field, i.e. multiple sources from the same field have the same shape.
\end{enumerate}

369 light curves were filtered and extracted by eye. These sources were further narrowed down by crossmatching with AAVSO's International Variable Star Index (VSX, \citealt{2006SASS...25...47W}), the Gaia DR3 Variable Catalog \citep{2023A&A...674A...1G}, the ATLAS variable catalog \citep{2018AJ....156..241H}, and searched if it was a known source in SIMBAD \citep{2000A&AS..143....9W}. We exclude a source from candidacy if it is a type of variable with high amplitudes and long periods.


\subsection{Search Results} \label{ss:search result}


\begin{figure*}[ht!]
    \centering
    \includegraphics[width=0.95\textwidth]{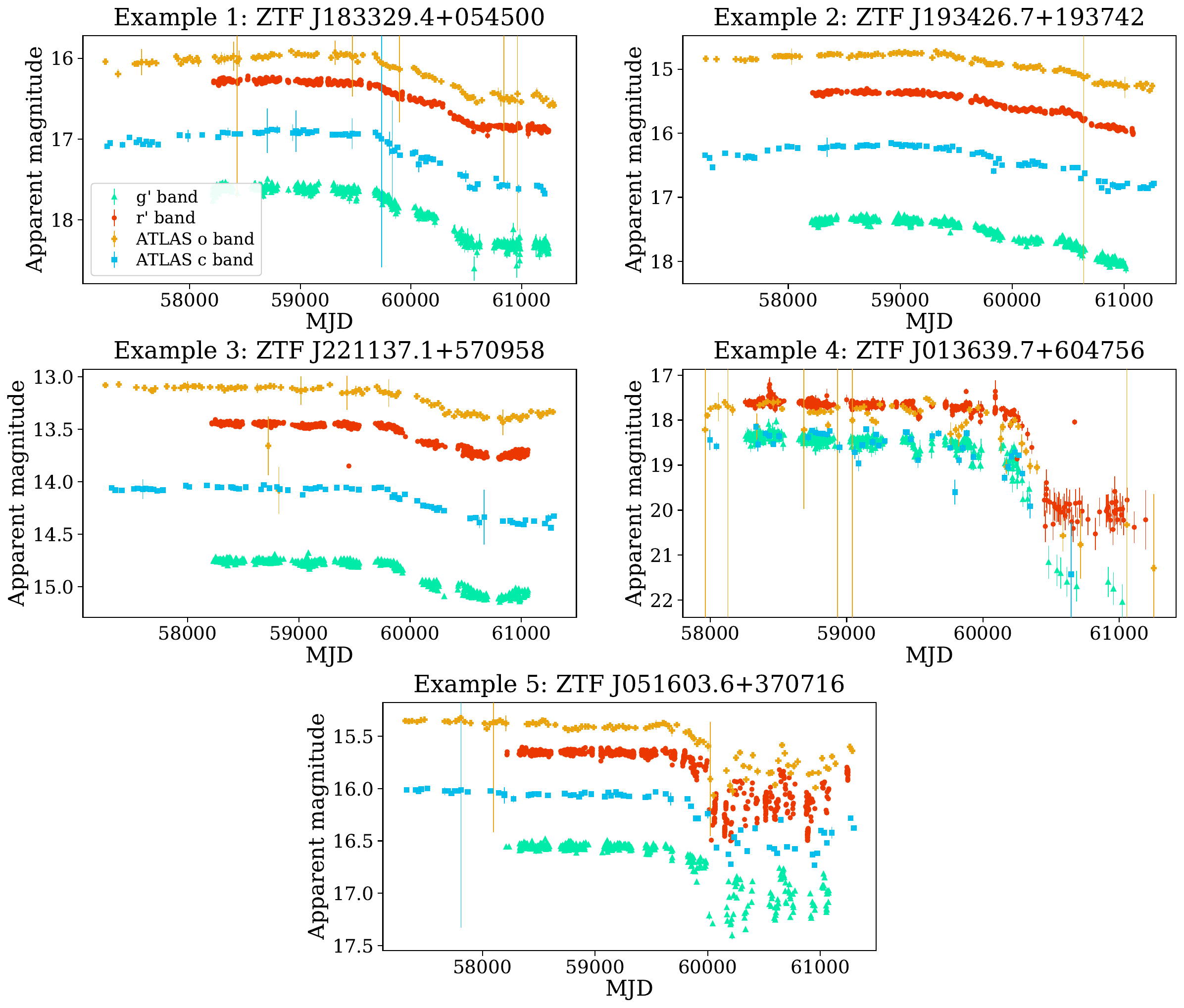}
    \caption{Light curve of 5 Example Faders, ZTF fluxes converted to magnitudes. (From top to bottom) Cyan and orange points are ATLAS $c$ and $o$ bands, and red and green for ZTF $r'$ and $g'$ bands.}
    \label{fig:candidate}
\end{figure*}

\begin{figure*}[h!]
    \centering
    \includegraphics[width=0.95\textwidth]{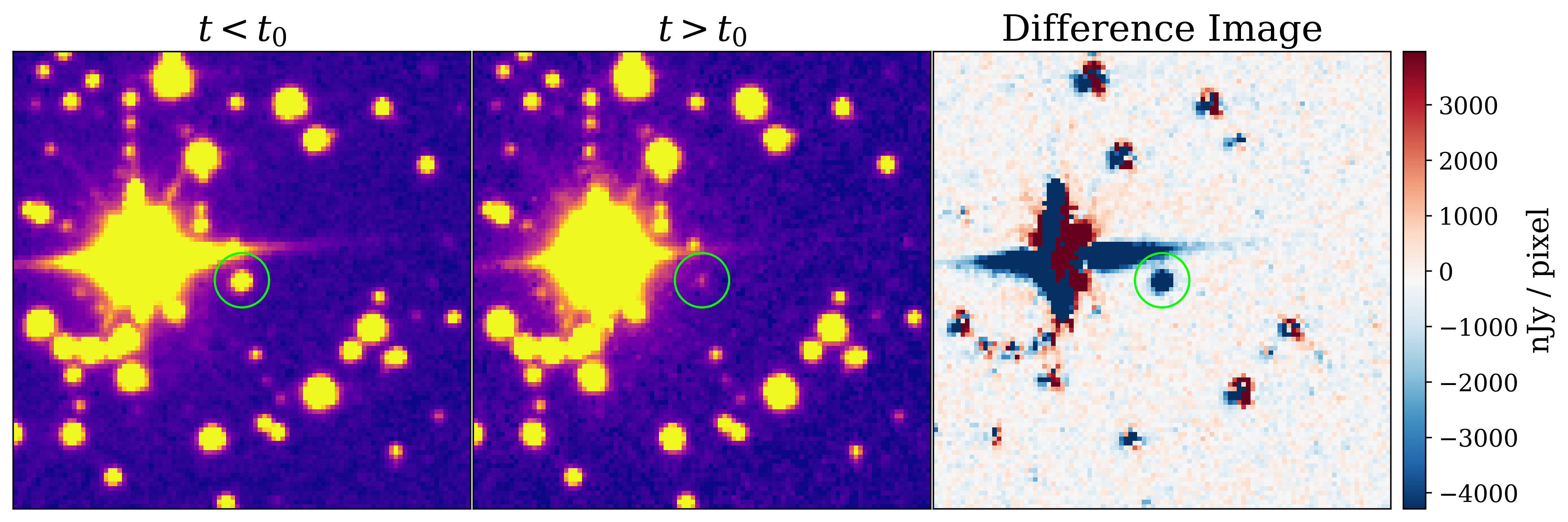}
    \caption{Median-stacked and Gaussian-smoothed $100"\times100"$ cutout images before and after the flux drop of Fader Example 4 (ZTF J013639.7+604756, \citealt{ztf_image_service}). The left panel shows 15 median-stacked images before the flux drop, the middle after the flux drop, and the right the direct difference image between the two.}
    \label{fig:cutout}
\end{figure*}

After cutting out the known variables, we are left with 125 ``faders" (see Figure \ref{fig:allcands} and Table \ref{tab:candidates}). All of these have a significant flux drop and roughly follow the flat-drop-flat curve shape. All 125 are candidates for follow-up observations and studies for spectra to determine the mechanism of the fade. We have selected 5 of the best examples within our findings and plotted them alongside their ATLAS \citep{2018PASP..130f4505T} light curve to show the flux drop is not an instrumental artifact (Figure \ref{fig:candidate}). Example 4 is particularly promising due to its 89\% and 95\% flux drop in g' and r' respectively and its flat bottom. This dramatic decrease in flux can be directly seen on ZTF images (Figure \ref{fig:cutout}), where the source visibly becomes fainter in the before and after images, with the difference image showing a clear negative footprint. 

\begin{figure*}[ht!]
    \centering
    \includegraphics[width=\linewidth]{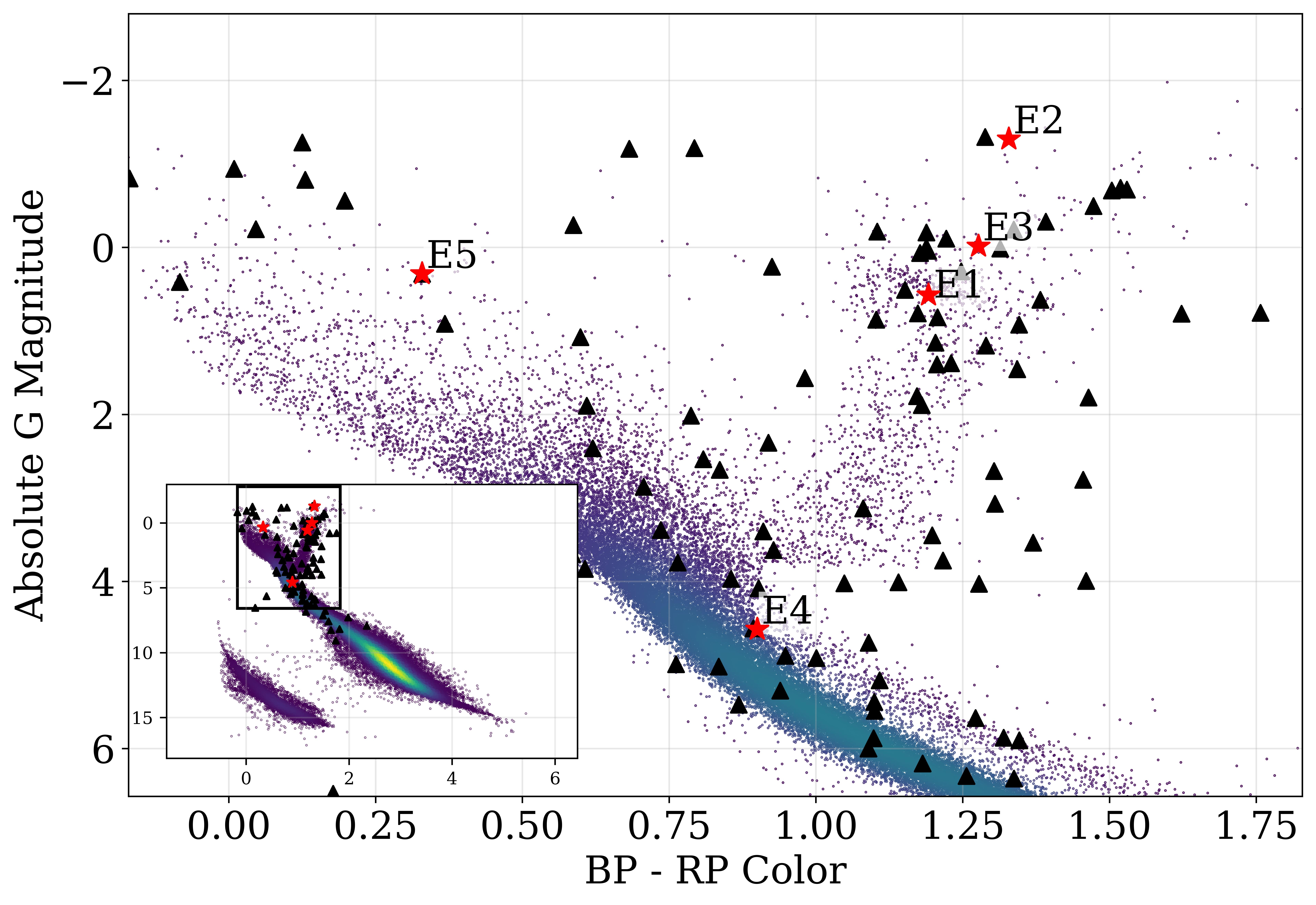}
    \caption{CMD of all faders. Blue-green points are Gaia-selected CMD stars, black points are all 125 faders, and the red points are the selected example faders. Most Faders are placed on the upper main sequence or the giant branch.}
    \label{fig:cmd}
\end{figure*}

The faders are placed on a color-magnitude diagram (CMD, from \citealt{2018A&A...616A..10G}) to classify them. Each fader's Gaia G band magnitude, BP-RP color, and parallax were queried from Gaia and corrected for extinction. Extinction is calculated by first calculating the $E(g-r)$ extinction from \citet{2019ApJ...887...93G}'s dust map, then converting the value to $E(BP-RP)$ and $A_G$ using Equations \ref{eq:egr} and \ref{eq:ag} respectively \citep{2019ApJ...887...93G, 2018A&A...616A...8A}. Most of the faders are either on the giant branch or upper main sequence (Figure \ref{fig:cmd}). FccSN progenitors are likely giant stars near the end of their life; the faders on the giant branch are more likely to be real candidates. Example 4 is in the main sequence, which is not where a candidate FccSN is expected to live, but is nonetheless an interesting source to follow up on. 

\begin{equation}
    E(BP-RP)=0.993(E(g-r))+.08
    \label{eq:egr}
\end{equation}

\begin{equation}
    A_G\sim2(E(BP-RP))
    \label{eq:ag}
\end{equation}

We can use ZTF's baseline of 7.5 years and the assumption that every FccSN is recovered by the search to calculate a statistical limiting rate of FccSNe within ZTF. This value can be interpolated through the ratio of massive stars visible in ZTF to the total number of massive stars in the Milky Way to calculate the galactic limiting rate (see Appendix \ref{app:massive star}). We can model the likelihood that a massive star undergoes FccSN as a Poisson distribution. With the assumption that none of the faders found in our search are real candidates, we can calculate that the observation rate of FccSN in ZTF is less than 61 per century at 99\% credibility, extrapolated to less than 147 per century in the Milky Way. In other words, if the actual rate is higher than this value, our ZTF survey would have a null result less than 1\% of the time. If, for example, we confirm that E4 is a real FccSN event, a rate range can be similarly calculated. The 99\% credibility rate range if the search finds one event is $1.4<r<99$ per century, extrapolated to $3.3<r<236$ for the entire Milky Way. The $1\sigma$ interval is $r=22^{+22}_{-13}$ and $r=52^{+52}_{-30}$ per century for ZTF and the Milky Way, respectively. The upper bounds on both calculations (null and one positive search result) are merely statistical limits and should not be taken as rate predictions. 



\section{Simulation and Predictions of FccSNe Rates}

In order to simulate and roughly estimate the Milky Way's FccSN rate, we combined a simple density profile and dust model to predict the formation rate and visibility of massive stars. 

\subsection{Density model} \label{ss:density}
We use a density distribution model as a proxy for the probability of stellar distribution and formation (see Section \ref{ss:formation}). The density model uses a combination of the disk model from \citet{2017MNRAS.465...76M} and the Hernquist profile \citep{1990ApJ...356..359H} for the bulge. The disk density profile is as follows:

\begin{equation}
    \rho_d(R, z)=\frac{\Sigma_{0}}{2z_{d}}\exp(\frac{-R}{R_{d}}-\frac{|z|}{z_{d}})
\end{equation}
where $\Sigma_0=887.0\ \rm M_{\odot}\ pc^{-2}$ is surface density, $z_d=76\ \rm pc$ scale height for younger massive stars \citep{quintana2025censusobstars1}, and $R_d=2600\ \rm pc$ scale length \citep{2017MNRAS.465...76M}. The Hernquist density profile for the bulge is as follows:

\begin{equation}
    \rho_b(r)=\frac{M_b}{2\pi}\frac{a}{r}\frac{1}{(a+r)^3}
\end{equation}
with the constants $M_b=1.3\times10^{10}\ \rm M_\odot$ total stellar bulge mass and $a=600\ \rm pc$ scale length \citep{2021arXiv211105466T}. The total stellar density profile thus becomes
\begin{equation}
    \rho=\rho_d+\rho_b.
\end{equation}
Note that these densities are of all matter in the galaxy and not just stellar densities. The bulge is less involved with stellar formation, with the majority of star formation in the disk \citep{Elia_2022}, but its removal does not significantly change the result of the simulation (${<}10\%$).

\subsection{Dust model}
\begin{figure*}[ht!]
    \centering
    \includegraphics[width=\linewidth]{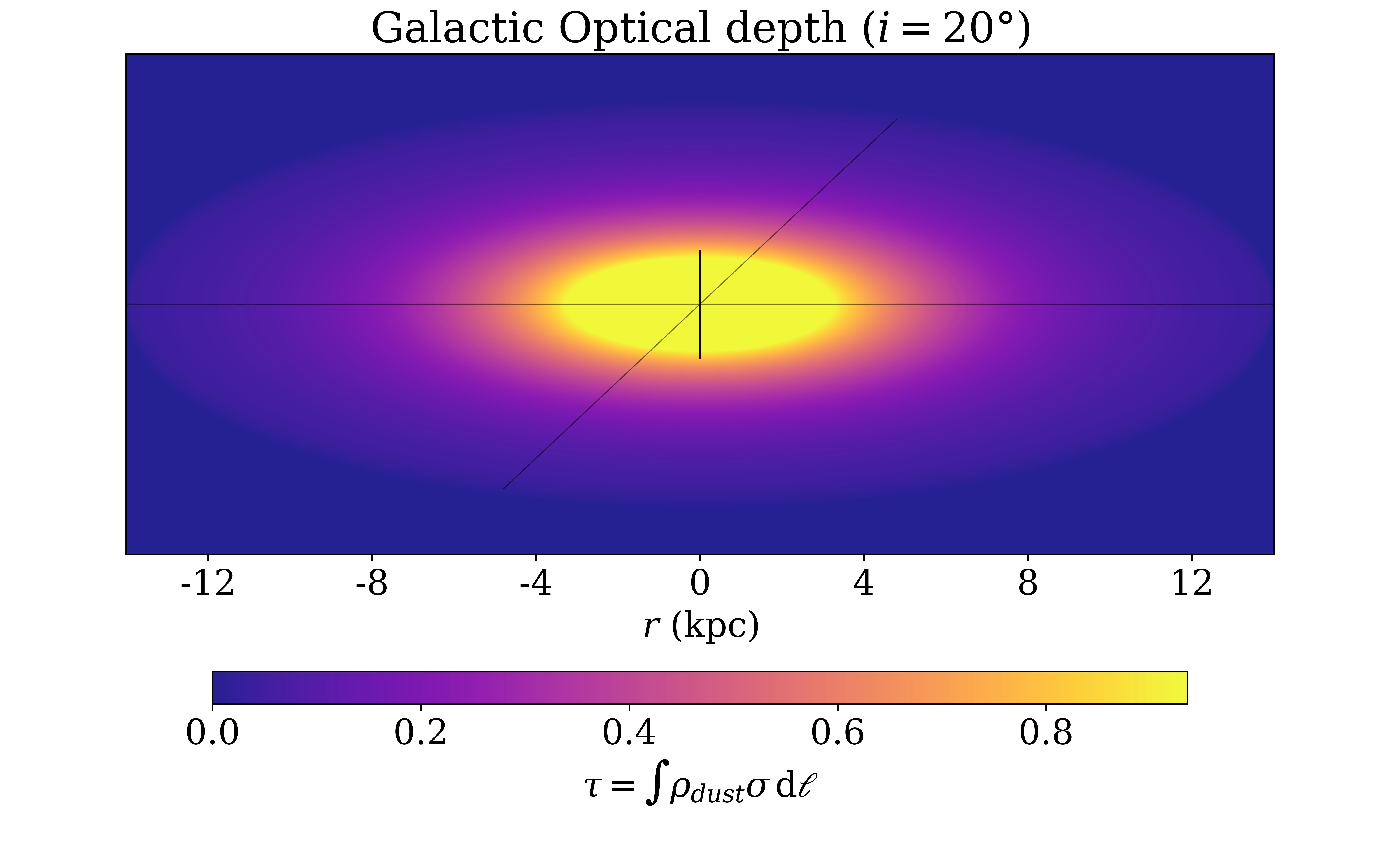}
    \caption{A visualization of the dust distribution model from an outside view and inclination $20\degree$. This represents the integrated line-of-sight optical depth at each point of the galaxy.}
    \label{fig:optdep}
\end{figure*}

The dust model primarily uses that of \citet{2018ApJ...858...75L}, an axisymmetric disk model parametrized as such:
\begin{equation}
    \rho_{\mathrm{dust}}(R, z)=\rho_{0, \mathrm{dust}}\exp(\frac{-R}{R_{\mathrm{dust}}}-\frac{|z|}{z_{\mathrm{dust}}})
\end{equation}
with $\rho_{0, \rm dust}$ the dust density at the Milky Way center, $R_{\rm dust}=3200\ \rm pc$ the scale length and $z_{\rm dust}=103\ \rm pc$ the scale height. Total extinction in magnitudes can thus be calculated by integrating the column density along the line of sight:
\begin{equation}
    A_{g'}=1.316\int^s_0\rho_{\mathrm{dust}}(s')\sigma_{\lambda}\ ds'.
\end{equation}
The base column density $\rho_{\mathrm{dust},0}\sigma_\lambda=0.0062\ \rm pc^{-1}$ will be used as a combined constant. The $1.316$ constant comes from the conversion of $A_V$ (used in \citealt{2018ApJ...858...75L}) and $A_{g'}$ (the filter used in our search). The dust model and the optical depth distribution is visualized in Figure \ref{fig:optdep}. 

Note that the dust model is a simplification; it does not take into account the irregularity of dust distribution, especially near the galactic center. For this paper, however, this model will suffice for a rough estimate of the FccSN rate.

\subsection{Star formation simulation} \label{ss:formation}

Because the objective of the density profile is to find the visibility of FccSNe in the Milky Way, the base coordinate system used in this simulation is a combination of distance of the sun from the galactic center and galactic plane ($R_0=8200\ \rm pc$, $z_0=15\ \rm pc$; \citealt{2019A&A...625L..10G, 2019A&A...632L...1S}) and celestial galactic longitude and latitude  $(s, l, b)$. Within this coordinate frame, the dust and density profiles were calculated by first converting to Cartesian coordinates, then to either cylindrical (disk density, dust) or spherical (bulge density) coordinates. In this model, the bound of the Milky Way were set to a radius of 30 kpc around the sun. With this bound, the simulated Milky Way was divided into $200^3$ segments, with radius and longitude each having 200 linearly spaced sections, and latitude having 200 logarithmically spaced sections to account for the bulk of density being near the galactic plane. 

\begin{figure*}[ht!]
    \centering
    \includegraphics[width=\linewidth]{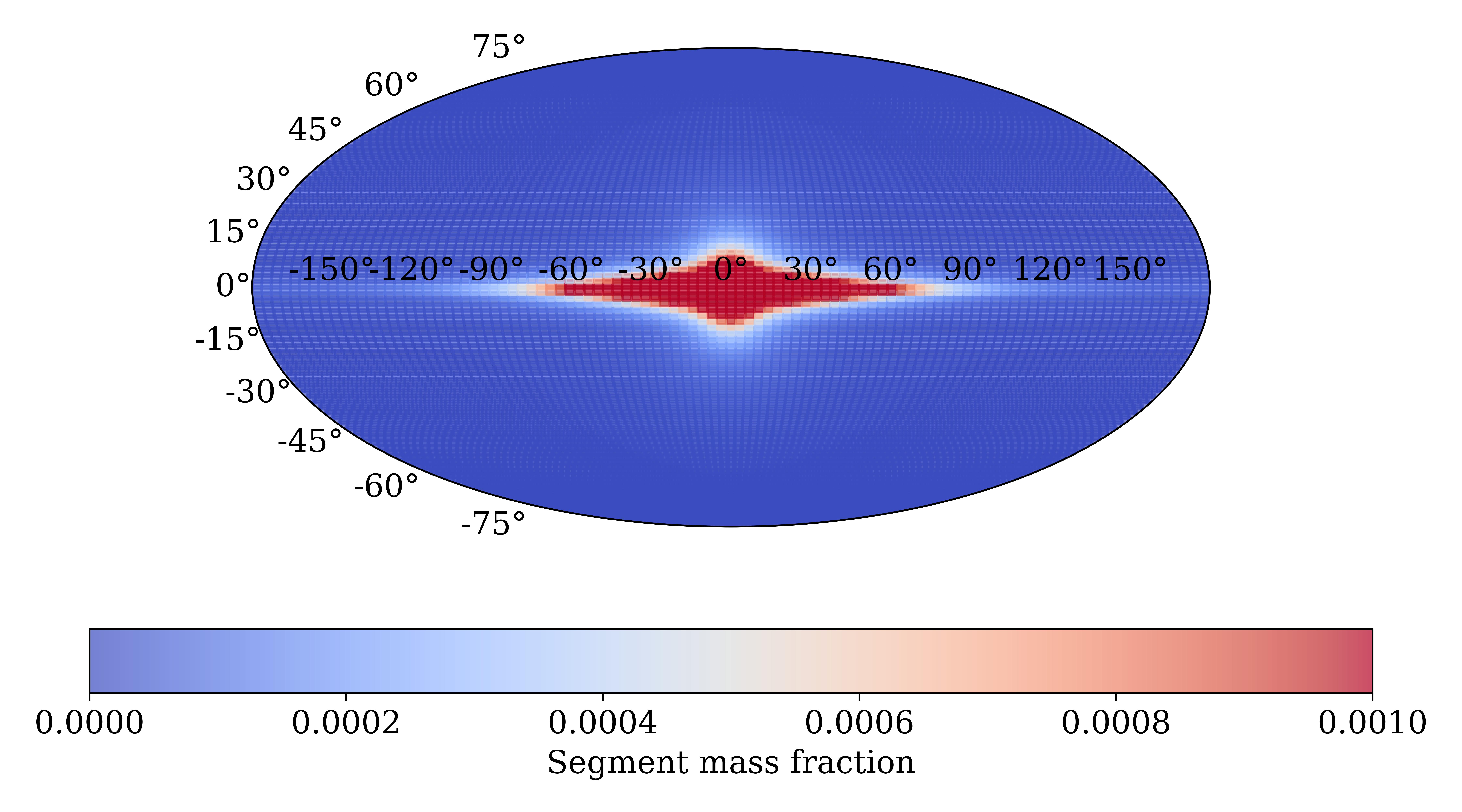}
    \caption{A visualization of the mass fraction of the Milky Way based on galactic coordinates. Each segment has a probability (expressed here by color, integrated along the line of sight) to form a star within a given simulation cycle.}
    \label{fig:galmass}
\end{figure*}

The proportion of total mass in each segment $M_{c}$ to the total galaxy mass $M_{tot}$ is used to calculate the probability of a star forming in that specific segment. In this exercise, all stars are isolated; the effects of binary systems are not factored in.
\begin{equation}
    \label{eq:9}
    M_{c}(s, l, b)=\rho(s, l, b)\times s^2\cos(b)\ \Delta s\ \Delta l\ \Delta b
\end{equation}
\begin{equation}
    \label{eq:10}
    M_{tot}=\int_0^{3\times10^4}\int_0^{2\pi}\int_{-\pi/2}^{\pi/2}\rho(s, l, b)\ s^2cos(b)\ db\ dl\ ds    
\end{equation}
\begin{equation}
    \label{eq:11}
    P_{\rm segment}=M_{c}/M_{tot}
\end{equation}
$\Delta s$, $\Delta l$, and $\Delta b$ are the spacing between the chosen coordinates. Figure \ref{fig:galmass} shows a visualization of the mass distribution with respect to galactic coordinates.

The probability distribution function of formed star mass is calculated using the Kroupa initial mass function (IMF; \citealt{2001MNRAS.322..231K}; see Appendix \ref{app:imf}). 
\begin{equation}
    p(M_{star})=\frac{\xi_{0, \alpha}M_{star}^{-\alpha}}{\int_{0.08}^{200}\xi_{0, \alpha}M^{-\alpha}dM}=\frac{M_{star}^{-\alpha}}{9.34}
\end{equation}
Simulated stars will fall under one of 200 geometrically spaced bins between 0.08 $\rm M_{\odot}$ and $200\  \rm M_{\odot}$ with probabilities corresponding to the cumulative distribution function (CDF)
\begin{equation}
    P(M_1<M<M_2)=\int_{M_1}^{M_2}\frac{M'^{-\alpha}}{9.34}\ dM'.
\end{equation}

A cycle of the simulation consists of continuously creating stars of masses corresponding to the CDF until the total mass of the created stars exceeds $165\ \rm M_{\odot}$, the Milky Way's star formation rate (SFR) per 100 years \citep{Licquia_2015}. The created stars are each placed in segments of the galaxy based on Equations \ref{eq:9}-\ref{eq:11}, where visibility and mass criteria are calculated (in Section \ref{ss:vc}). This process is repeated for $10^4$ cycles, where the mean number of visible massive stars per cycle is be calculated.

The current star formation rate is used to calculate the number of FccSNe because their progenitors are massive stars (${>}20\ M_{\odot}$) born at most $10^7$ years ago, based on their main sequence lifetime. It can be assumed that the stellar formation rate was similar at that time. The formation of a star within a year is roughly one-to-one with the death of the same star. Note that we assume the distribution of stellar formation is equivalent to the density distribution.

\subsection{Visibility and mass criteria} \label{ss:vc}
Each created star undergoes a filter to check if they are massive enough to undergo FccSN and if they are bright enough to be seen by certain ground-based optical telescopes. If a simulated star exceeds 20 $\rm M_{\odot}$, we proceed to make a giant phase magnitude calculation. Since FccSN progenitors are giant phase stars at the end of their life, we model stars of a range of masses in MESA \citep{2011ApJS..192....3P} and use the giant phase luminosity of the modeled stars for this simulation. This luminosity is converted into absolute bolometric magnitude, then finally a bolometric correction and extinction model is applied to convert to r' band apparent magnitude \citep{2018ApJS..236...47W}. If the simulated star's apparent magnitude is brighter than 21 mag, the star is added to the count of visible massive stars. Finally, if the telescope of interest has a limited field (e.g. in the northern hemisphere), only stars in visible coordinates will be counted. 


\subsection{Simulation results} \label{ss:simres}
The simulation, without any telescope field visibility constraints, yields $0.538\pm0.011$ stars over 20 $\rm M_\odot$ formed per 100 years. With the ZTF field and magnitude visibility constraints, the simulation formed $0.234\pm0.006$ per century. This value is equivalent to an event every 185 and 430 years, respectively. Applying the same IMF calculation to stars ${\geq} 10\ \rm M_\odot$ yields an overall ccSN rate of $1.45$ per century. The resulting 37\% FccSN rate directly follows the Kroupa IMF with the assumed $M\geq20\ \rm M_\odot$ threshold, and is much greater than the 20\% from \citet{2016ApJ...821...38S}. This discrepancy may be due to the $M\geq20\rm M_\odot$ assumption (and not considering the ``islands of explodability") or a higher effective mass threshold. A 20\% fraction corresponds to a mass threshold of $M\geq31 \rm \ M_\odot$.

\subsection{Predictions for LSST and Roman}

The limitation of this search is ZTF's angular resolution and field depth. Using ZTF only allows for a search within the Milky Way and a shallow search of nearby galaxies; the total number of visibly forming (and thus dying) stars is limited by the observable volume. Fortunately, new and upgraded time-domain surveys are coming online in the forms of Vera C. Rubin Observatory's LSST \citep{2019ApJ...873..111I} and Roman's WFI \citep{2024SPIE13092E..0SS}. ZTF's angular resolution and greatest limiting magnitude is 2.0" and 20.5 mag respectively, but LSST ($\theta=0.7"$, $\rm mag_{\mathrm{lim}}\approx23$) and the survey conducted by WFI ($\theta=0.1"$, $\rm mag_{\mathrm{lim}}\approx25$) can expand the search to much further galaxies, opening up the total star formation rate for hopefully more frequent events. Furthermore, Roman has the added advantage of observing in the IR, where dust obscuration does not lead to a significant flux drop, reducing the likelihood of encountering false positives that would otherwise be found with an optical search such as the one described in this paper.

A naive calculation of the occurrence of such events was done using the instruments' limits. For a lower limit of the possible number of FccSN, we use the angular resolution of each instrument and the average physical separation between two stars within a galaxy to calculate the local volume in which individual stars can be resolved. We use a toy assumption of ${\sim}1$ pc average separation between individual stars, using separation in the local solar volume \citep{2023A&A...670A..19G} since galactic centers will be too crowded for resolution, and the outer disk/halo will be too sparse to be able to observe an adequate survey. LSST will be able to resolve stars 1 pc apart up to 0.3 Mpc, and WFI up to 2 Mpc. We then convolve the IMF of massive stars (${\geq}20\ \rm M_\odot$) with the SFRs of the galaxies with significant star formation within these distances to calculate the rough estimate of the number of FccSNe the instruments can resolve. LSST's visible local volume includes both the Milky Way ($SFR=1.65\ \rm M_\odot$; \citealt{Licquia_2015}) and the Large Magellanic Cloud ($SFR=0.2\ \rm M_\odot$; \citealt{2009AJ....138.1243H}), and Roman adds M31 ($SFR=0.35\ \rm M_\odot$; \citealt{2016MNRAS.456.4128R}) and M33 ($SFR=0.45$; \citealt{2009A&A...493..453V}). From this, we estimate LSST and Roman can resolve $0.46$ and $0.65$ FccSNe per century (see Appendix \ref{app:predictions}).

For an upper limit, we do a similar calculation using the magnitude limit of each instrument. This method of searching will ignore resolvability of individual disappearing sources, and rather search for flux drops of galaxies within the local universe. For the visibility criteria, we set an absolute magnitude threshold on massive stars $M\geq20M_\odot$, obtained from MESA models \citep{2011ApJS..192....3P}. Within the calculated distances, SFR densities (SFRD) of $0.019\ \mathrm{M_\odot\ Mpc^{-3}\ yr^{-1}}; d=10\  \mathrm{Mpc}$ \citep{Karachentsev_2010} and $0.0079\ \mathrm{M_\odot\ Mpc^{-3}\ yr^{-1}}; d=75\ \mathrm{Mpc}$ \citep{2021A&A...650A..68V}, with a cutoff between the two values at $40\ \rm Mpc$, were used for the local universe. From this, we obtain a FccSN rate of $7.2\ \mathrm{yr^{-1}}$ and $100\ \mathrm{yr^{-1}}$ for LSST and Roman, respectively (see Appendix \ref{app:predictions}). These numbers are the maximum possible number of events observable in the most ideal circumstances: constant observation of all star-forming galaxies within the magnitude-limit distances, assuming they are all inclined face-on. The actual observed number purely searching for flux drops in galaxies within the local universe will likely be significantly fewer than this. 

\section{Discussion} \label{sec:discussion}
The 125 faders we found will require further study to trial their candidacies. Infrared follow-ups and spectra will be crucial to distinguish an actual FccSN from a dust obscuration. A steady IR flux concurrent with decreasing optical flux would point to dues obscuration, given the reduced opacity of dust to IR radiation. Continued monitoring of these sources is also crucial to rule out the possibility that the observation on hand is only an ingress of a long eclipse (e.g. \citealt{For_s_Toribio_2025}) or a long-period variable. The odds that there is a real FccSN event in the faders are low, assuming our prediction from the simulation is accurate; ZTF's total observing period is roughly 7.5 years, and the simulation expects to see an event every 430 years in the ZTF footprint. However, all faders are interesting sources to follow up on, given their light curve shape and their placements on the CMD (Figure \ref{fig:cmd}) being mostly within the giant branch. 

In particular, Example 4 (ZTF J013639.7+604756) is a promising potential FccSN candidate due to its light curve shape, a near 2 magnitude drop in both ZTF bands, and the clear flat-drop-flat shape. Its placement on the CMD (Figure \ref{fig:cmd}) can suggest otherwise; its placement on the main sequence (and spectral class G, estimated from its BP-RP color) would dissuade the direct collapse conclusion. However, the error bar on its Gaia parallax is fairly large (about 26\%), so it is possible that E4 is brighter than its current placement on the CMD. There are a few other possible explanations for E4's dramatic flux drop. One specific explanation is R Coronae Borealis (R CrB) stars; their fast, sudden drops in flux last anywhere between 10 to $>$1000 days \citep{2024MNRAS.527.9274S}. However, the typical timescale of the flux drop is less than 100 days, much shorter than E4's drop time of $>$500 days. The most likely conclusion is obscuration or eclipse by optically thick dust, which can explain the achromatic (in the optical) behavior of the light curve. Infrared observations and spectra (neither of which could be found for E4) could rule out these possibilities. If the spectrum appears blueshifted, it can be a rapidly expanding dust shell coming from a FccSN \citep{2026ApJ...999L..25D}. Confirmation of E4 as a real candidate will allow for constraints on the population of these events. 


The rate obtained from the simulation ($0.789\pm0.022$ per century in the Milky Way) is significantly less than the statistical limiting rate calculated from the null result of the ZTF search (147 per century). Given the small rate obtained from the simulation, this difference is expected; the null result of the search cannot constrain the models yet. Additionally, the ccSN failure rate of 37\% obtained from the simulation is much higher than what \citet{2014ApJ...785...28K} predicts (20\%), so there is a possibility that the predicted rate is also an overestimation. The assumptions made would tend towards an overestimation, especially when using $M{\geq}20\ \rm M_\odot$ as an absolute threshold for a ccSN failure. However, with a deeper field and longer observation time, constraining the rate and understanding its population should become easier. Finding a real candidate in a search like this work would greatly narrow the limiting rate range.


For our simulations and predictions, we chose $M>20\rm M_\odot$ to be consistent with \citet{2016ApJ...821...38S}'s models predicting the likelihood of FccSNe. However, we did not account for the possibility of a lower mass progenitor directly collapsing, as we see in the case of \citet{2026ApJ...999L..25D} with a $M=12{\sim}13\rm M_\odot$ progenitor. It is difficult to consider these theoretical ``exceptions" due to the lack of both our understanding and observations of such events to form a empirical limiting rate on FccSNe. Using LSST and Roman to survey for such events might unlock population of these events. However, such a search will require extensive galaxy monitoring by these instruments, such as the one described in \citet{2008ApJ...684.1336K}. Such a search may be aided by single JWST images of each star-forming galaxies to set baseline fluxes. Building a known population of FccSNe can constrain their physical mechanisms, accelerating our understanding of stellar evolution.

\begin{software}
    Astropy \citep{astropy2022}, Numpy \citep{harris2020numpy}, Matplotlib \citep{hunter2007matplotlib}, Pandas \citep{pandas_dev}, Scipy \citep{virtanen2020scipy}, Claude \citep{anthropic_claude}, MESA \citep{2011ApJS..192....3P}
\end{software}

\begin{acknowledgments}
The authors thank Kishalay De for helpful discussion during the development of this manuscript. 
A.H. is supported by the National Science Foundation Graduate Research Fellowship Program. 
\end{acknowledgments}

\bibliography{ref}
\bibliographystyle{aasjournalv7.1}

\appendix

\section{Faders from ZTF Search} \label{app:candidates}
\startlongtable

\begin{deluxetable}{ccccccc}
    \tablecaption{All of the found faders. Right ascension and declination are in degrees, $\Delta F/F_0$ are flux fractional fades, and $\Delta t$ are in days. Flux and timescale values are taken directly from the linear model parameters (see Equation \ref{eq:16}).} 
    \label{tab:candidates}
    \tablehead{
    \colhead{Source} & \colhead{RA} & \colhead{Dec} & \colhead{$\Delta F_{g'}/F_{g', 0}$} & \colhead{$\Delta F_{r'}/F_{r', 0}$} & \colhead{$\Delta t_{g'} \ (d)$} & \colhead{$\Delta t_{r'}\;(d)$}
    }
    \startdata
        ZTF J063441.1-252323 & 98.6714 & -25.3898 & 0.4 & 0.32 & 2300 & 2000 \\ 
        ZTF J062238.4-243117 & 95.6598 & -24.5215 & 0.19 & 0.33 & 1200 & 1400 \\ 
        ZTF J074647.3-261829 & 116.6969 & -26.3083 & 0.21 & 0.2 & 500 & 1000 \\ 
        ZTF J120827.4-262309 & 182.1142 & -26.386 & 0.21 & 0.26 & 2400 & 2700 \\ 
        ZTF J165425.7-271514 & 253.6072 & -27.2541 & 0.39 & 0.33 & 1200 & 1400 \\ 
        ZTF J172133.0-223916 & 260.3874 & -22.6547 & 0.86 & 0.57 & 1300 & 1200 \\ 
        ZTF J172404.6-210948 & 261.0192 & -21.1635 & 1.07\tablenotemark{1} & 0.91 & 1900 & 2300 \\ 
        ZTF J175819.7-224637 & 269.5822 & -22.777 & 0.41 & 0.24 & 2900 & 2500 \\ 
        ZTF J185112.8-242204 & 282.8034 & -24.3678 & 0.13 & 0.12 & 2800 & 2700 \\ 
        ZTF J184842.0-235544 & 282.1748 & -23.9291 & 0.83 & 0.5 & 1900 & 2600 \\ 
        ZTF J190131.5-265704 & 285.3812 & -26.9513 & 0.59 & 0.6 & 800 & 2200 \\ 
        ZTF J185611.0-205748 & 284.046 & -20.9636 & 0.3 & 0.35 & 2900 & 2400 \\ 
        ZTF J055058.5-182227 & 87.7439 & -18.3744 & 0.08 & 0.05 & 1500 & 1400 \\ 
        ZTF J090907.7-195457 & 137.2821 & -19.916 & 0.46 & 0.65 & 1400 & 1500 \\ 
        ZTF J101032.6-151833 & 152.6357 & -15.3092 & 0.24 & 0.11 & 2500 & 2700 \\ 
        ZTF J143510.7-134730 & 218.7944 & -13.7917 & 0.22 & 0.18 & 2900 & 2800 \\ 
        ZTF J164836.6-181939 & 252.1523 & -18.3275 & 1.11\tablenotemark{1} & 0.39 & 1900 & 900 \\ 
        ZTF J170437.9-144626 & 256.1579 & -14.7741 & 0.68 & 0.6 & 2200 & 2000 \\ 
        ZTF J171851.7-174651 & 259.7153 & -17.781 & 0.62 & 0.47 & 1200 & 1400 \\ 
        ZTF J172044.5-184911 & 260.1855 & -18.8198 & 0.33 & 0.3 & 1200 & 1000 \\ 
        ZTF J180414.7-202116 & 271.0613 & -20.3546 & 0.38 & 0.26 & 2700 & 2700 \\ 
        ZTF J190707.0-202809 & 286.7793 & -20.4694 & 0.6 & 0.52 & 1800 & 1800 \\ 
        ZTF J192226.0-193815 & 290.6085 & -19.6376 & 0.46 & 0.39 & 1700 & 1800 \\ 
        ZTF J192345.4-133423 & 290.9391 & -13.5732 & 0.37 & 0.33 & 2200 & 2200 \\ 
        ZTF J211803.3-171252 & 319.5137 & -17.2145 & 1.58\tablenotemark{1} & 0.47 & 2900 & 1000 \\ 
        ZTF J061220.6-080618 & 93.086 & -8.105 & 0.47 & 0.34 & 1500 & 1900 \\ 
        ZTF J063740.5-115521 & 99.4186 & -11.9227 & 0.43 & 0.35 & 1300 & 1100 \\ 
        ZTF J062828.1-110726 & 97.1172 & -11.1239 & 0.75 & 0.43 & 900 & 1000 \\ 
        ZTF J070232.1-121336 & 105.6338 & -12.2267 & 0.5 & 0.37 & 1400 & 1200 \\ 
        ZTF J065912.7-115841 & 104.803 & -11.9783 & 0.36 & 0.33 & 2600 & 2400 \\ 
        ZTF J073419.8-082820 & 113.5825 & -8.4724 & 0.63 & 0.47 & 1700 & 1800 \\ 
        ZTF J113212.4-082746 & 173.0517 & -8.4629 & 0.39 & 0.35 & 2100 & 2000 \\ 
        ZTF J143041.6-070755 & 217.6732 & -7.1322 & 0.38 & 0.31 & 1500 & 1600 \\ 
        ZTF J174635.6-125824 & 266.6483 & -12.9735 & 0.73 & 0.71 & 1600 & 1400 \\ 
        ZTF J190859.3-092140 & 287.247 & -9.3613 & 0.45 & 0.31 & 1300 & 1200 \\ 
        ZTF J065451.4-010643 & 103.7142 & -1.112 & 0.36 & 0.16 & 0 & 100 \\ 
        ZTF J072752.2-021001 & 111.9676 & -2.1672 & 0.33 & 0.34 & 1600 & 2200 \\ 
        ZTF J092947.7-055357 & 142.4489 & -5.8993 & 0.35 & 0.42 & 1600 & 1900 \\ 
        ZTF J183949.0-053810 & 279.954 & -5.6362 & 0.15 & 0.14 & 300 & 300 \\ 
        ZTF J184114.5-025157 & 280.3105 & -2.8659 & 0.43 & 0.35 & 1400 & 1700 \\ 
        ZTF J184331.2-041230 & 280.8799 & -4.2086 & 0.3 & 0.23 & 1900 & 1800 \\ 
        ZTF J190415.7-013722 & 286.0655 & -1.6228 & 0.48 & 0.39 & 1100 & 1100 \\ 
        ZTF J190536.8-001523 & 286.4033 & -0.2564 & 0.54 & 0.38 & 1900 & 1500 \\ 
        ZTF J191502.0-010158 & 288.7584 & -1.0328 & 0.38 & 0.3 & 1500 & 1000 \\ 
        ZTF J192724.1-005137 & 291.8506 & 0.8604 & 0.39 & 0.32 & 1500 & 1400 \\ 
        ZTF J194540.8-050934 & 296.42 & -5.1595 & 0.5 & 0.44 & 2300 & 1800 \\ 
        ZTF J021521.5+064223 & 33.8394 & 6.7064 & 0.44 & 0.35 & 1000 & 1000 \\ 
        ZTF J064727.2+031231 & 101.8635 & 3.2088 & 0.92 & 0.76 & 900 & 1300 \\ 
        ZTF J144401.7+033427 & 221.007 & 3.5743 & 0.44 & 0.32 & 1700 & 1100 \\ 
        ZTF J175509.3+010750 & 268.7888 & 1.1306 & 0.53 & 0.4 & 1200 & 1100 \\ 
        ZTF J182232.4+040643 & 275.6349 & 4.112 & 0.2 & 0.16 & 1300 & 1600 \\ 
        ZTF J183329.4+054500 & 278.3727 & 5.7502 & 0.67 & 0.51 & 1000 & 1100 \\ 
        ZTF J194957.6+073140 & 297.4898 & 7.5278 & 0.58 & 0.5 & 1500 & 1900 \\ 
        ZTF J200650.7+012544 & 301.7112 & 1.4291 & 0.47 & 0.38 & 2000 & 2000 \\ 
        ZTF J034000.8+132416 & 55.0035 & 13.4045 & 0.52 & 0.33 & 1100 & 1300 \\ 
        ZTF J061256.1+151534 & 93.2337 & 15.2595 & 0.35 & 0.25 & 500 & 500 \\ 
        ZTF J062718.7+150523 & 96.8281 & 15.09 & 0.71 & 0.24 & 600 & 100 \\ 
        ZTF J191033.4+105934 & 287.6391 & 10.9928 & 0.21 & 0.29 & 600 & 1600 \\ 
        ZTF J185447.8+110359 & 283.699 & 11.0664 & 0.63 & 0.54 & 1600 & 1400 \\ 
        ZTF J193931.4+140626 & 294.881 & 14.1075 & 0.61 & 0.58 & 900 & 2300 \\ 
        ZTF J191631.9+152001 & 289.1328 & 15.3337 & 0.75 & 0.49 & 1200 & 1200 \\ 
        ZTF J195827.3+150617 & 299.6137 & 15.1049 & 0.42 & 0.41 & 1200 & 1200 \\ 
        ZTF J195455.8+150520 & 298.7323 & 15.0889 & 0.51 & 0.43 & 1500 & 1400 \\ 
        ZTF J201051.2+102055 & 302.7135 & 10.3488 & 0.61 & 0.37 & 900 & 900 \\ 
        ZTF J201242.1+132824 & 303.1754 & 13.4734 & 0.65 & 0.29 & 1700 & 1600 \\ 
        ZTF J201345.5+122011 & 303.4397 & 12.3364 & 0.24 & 0.2 & 2300 & 2200 \\ 
        ZTF J201116.2+135436 & 302.8173 & 13.9102 & 0.96 & 0.49 & 2000 & 1700 \\ 
        ZTF J204810.8+120406 & 312.045 & 12.0686 & 0.35 & 0.39 & 2500 & 2400 \\ 
        ZTF J204207.2+132149 & 310.5301 & 13.3637 & 0.54 & 0.48 & 1700 & 1500 \\ 
        ZTF J050426.9+154908 & 76.112 & 15.819 & 0.72 & 0.71 & 1800 & 1700 \\ 
        ZTF J052009.0+211056 & 80.0374 & 21.1824 & 0.38 & 0.32 & 2100 & 2200 \\ 
        ZTF J194531.8+162732 & 296.3825 & 16.4591 & 0.66 & 0.51 & 2000 & 1700 \\ 
        ZTF J194531.6+162732 & 296.3815 & 16.4591 & 0.24 & 0.32 & 2100 & 2600 \\ 
        ZTF J195026.2+194948 & 297.609 & 19.8302 & 0.47 & 0.3 & 2400 & 1000 \\ 
        ZTF J193426.7+193742 & 293.6112 & 19.6285 & 0.72 & 0.51 & 2100 & 1900 \\ 
        ZTF J213221.8+155114 & 323.091 & 15.854 & 0.34 & 0.33 & 1600 & 1600 \\ 
        ZTF J231034.2+184202 & 347.6424 & 18.7006 & 0.39 & 0.28 & 1000 & 1300 \\ 
        ZTF J024334.6+264005 & 40.8943 & 26.6683 & 0.48 & 0.34 & 500 & 600 \\ 
        ZTF J055012.6+252419 & 87.5527 & 25.4053 & 0.18 & 1.74\tablenotemark{1} & 1800 & 1700 \\ 
        ZTF J190510.4+283021 & 286.2932 & 28.5059 & 0.6 & 0.37 & 1300 & 1000 \\ 
        ZTF J192606.5+232316 & 291.5269 & 23.388 & 0.15 & 0.11 & 2300 & 1800 \\ 
        ZTF J193518.2+253836 & 293.826 & 25.6436 & 0.94 & 0.78 & 1200 & 1100 \\ 
        ZTF J195943.2+224444 & 299.9298 & 22.7457 & 0.56 & 0.26 & 1200 & 1100 \\ 
        ZTF J195943.1+224444 & 299.9295 & 22.7457 & 0.39 & 0.15 & 1200 & 1100 \\ 
        ZTF J195021.4+244632 & 297.589 & 24.7757 & 0.48 & 0.41 & 1000 & 1000 \\ 
        ZTF J195923.0+272240 & 299.8458 & 27.3779 & 0.57 & 0.52 & 2000 & 1900 \\ 
        ZTF J195814.7+274300 & 299.5611 & 27.7169 & 0.67 & 0.46 & 1900 & 1200 \\ 
        ZTF J213500.5+235225 & 323.7522 & 23.8737 & 0.17 & 0.11 & 1700 & 1600 \\ 
        ZTF J235802.8+325909 & 359.5115 & 32.986 & 0.55 & 0.38 & 1800 & 1500 \\ 
        ZTF J195833.0+330937 & 299.6377 & 33.1604 & 0.59 & 0.29 & 1300 & 1100 \\ 
        ZTF J195340.3+313509 & 298.4181 & 31.5859 & 0.58 & 0.4 & 1900 & 2000 \\ 
        ZTF J194052.6+330417 & 295.219 & 33.0714 & 0.67 & 0.4 & 2800 & 2300 \\ 
        ZTF J203001.8+362610 & 307.5076 & 36.4362 & 0.43 & 0.3 & 700 & 800 \\ 
        ZTF J235935.4+425536 & 359.8975 & 42.9267 & 0.24 & 0.45 & 1700 & 1900 \\ 
        ZTF J003104.1+364034 & 7.767 & 36.6762 & 0.27 & 0.42 & 2100 & 2100 \\ 
        ZTF J195220.1+371645 & 298.0836 & 37.2793 & 0.16 & 0.08 & 1800 & 1700 \\ 
        ZTF J203259.8+411419 & 308.249 & 41.2388 & 0.57 & 0.51 & 1400 & 1400 \\ 
        ZTF J210704.2+390021 & 316.7674 & 39.006 & 0.48 & 0.44 & 1000 & 600 \\ 
        ZTF J021114.5+492702 & 32.8105 & 49.4508 & 0.41 & 0.38 & 1500 & 1400 \\ 
        ZTF J171435.4+470850 & 258.6477 & 47.1473 & 0.61 & 0.38 & 2200 & 1700 \\ 
        ZTF J190517.6+450702 & 286.3234 & 45.1175 & 0.4 & 0.39 & 1100 & 1100 \\ 
        ZTF J192950.0+460951 & 292.4584 & 46.1643 & 0.3 & 0.31 & 1800 & 2200 \\ 
        ZTF J195624.9+491458 & 299.1036 & 49.2497 & 0.41 & 0.3 & 1600 & 1600 \\ 
        ZTF J202551.2+501753 & 306.4634 & 50.2981 & 0.36 & 0.34 & 2400 & 2500 \\ 
        ZTF J211830.8+440531 & 319.6284 & 44.0922 & 0.42 & 0.38 & 1300 & 1600 \\ 
        ZTF J211058.4+465724 & 317.7435 & 46.9569 & 0.56 & 0.5 & 1700 & 1700 \\ 
        ZTF J214316.6+485828 & 325.8191 & 48.9745 & 0.23 & 0.14 & 200 & 100 \\ 
        ZTF J024437.2+522229 & 41.155 & 52.3749 & 0.31 & 0.15 & 1900 & 1000 \\ 
        ZTF J044907.6+524754 & 72.2817 & 52.7984 & 0.51 & 0.5 & 1900 & 2100 \\ 
        ZTF J221406.1+521126 & 333.5255 & 52.1906 & 0.56 & 0.8 & 1500 & 1600 \\ 
        ZTF J221311.2+542531 & 333.2966 & 54.4254 & 0.44 & 0.33 & 1400 & 1400 \\ 
        ZTF J221137.1+570958 & 332.9047 & 57.1663 & 0.34 & 0.3 & 600 & 800 \\ 
        ZTF J225725.6+513006 & 344.3567 & 51.5017 & 0.32 & 0.22 & 2800 & 2000 \\ 
        ZTF J003732.1+625113 & 9.3839 & 62.8538 & 0.11 & 0.11 & 300 & 200 \\ 
        ZTF J013639.7+604756 & 24.1653 & 60.7991 & 1.39\tablenotemark{1} & 2.25\tablenotemark{1} & 700 & 2100 \\ 
        ZTF J045857.8+680636 & 74.7408 & 68.1101 & 0.49 & 0.35 & 700 & 700 \\ 
        ZTF J015910.0+861139 & 29.7915 & 86.1942 & 0.17 & 0.17 & 1200 & 1500 \\ 
        ZTF J074250.0+051100 & 115.7082 & 5.1835 & 0.5 & 0.42 & 1600 & 1700 \\ 
        ZTF J190949.9+173959 & 287.4579 & 17.6664 & 0.51 & 0.66 & 700 & 1000 \\ 
        ZTF J194926.3+141553 & 297.3594 & 14.2648 & 0.5 & 0.37 & 1200 & 1200 \\ 
        ZTF J194126.6+323948 & 295.3607 & 32.6635 & 0.52 & 0.46 & 1100 & 1000 \\ 
        ZTF J051603.6+370716 & 79.0149 & 37.1212 & 0.55 & 0.61 & 500 & 300 \\ 
        ZTF J203654.2+375548 & 309.2257 & 37.9302 & 0.96 & 0.81 & 2000 & 1900 \\ 
        ZTF J012626.1+502622 & 21.6086 & 50.4397 & 0.49 & 0.45 & 2500 & 2700 \\ 
    \enddata
    \tablenotetext{1}{$\Delta F/F_0$ values can be greater than 1 due to source blending or reference flux measurements above a source's baseline, making the fitted model have a negative $F_0-\Delta F$ value.}
\end{deluxetable}

\begin{figure*}
    \centering
    \includegraphics[width=0.95\textwidth]{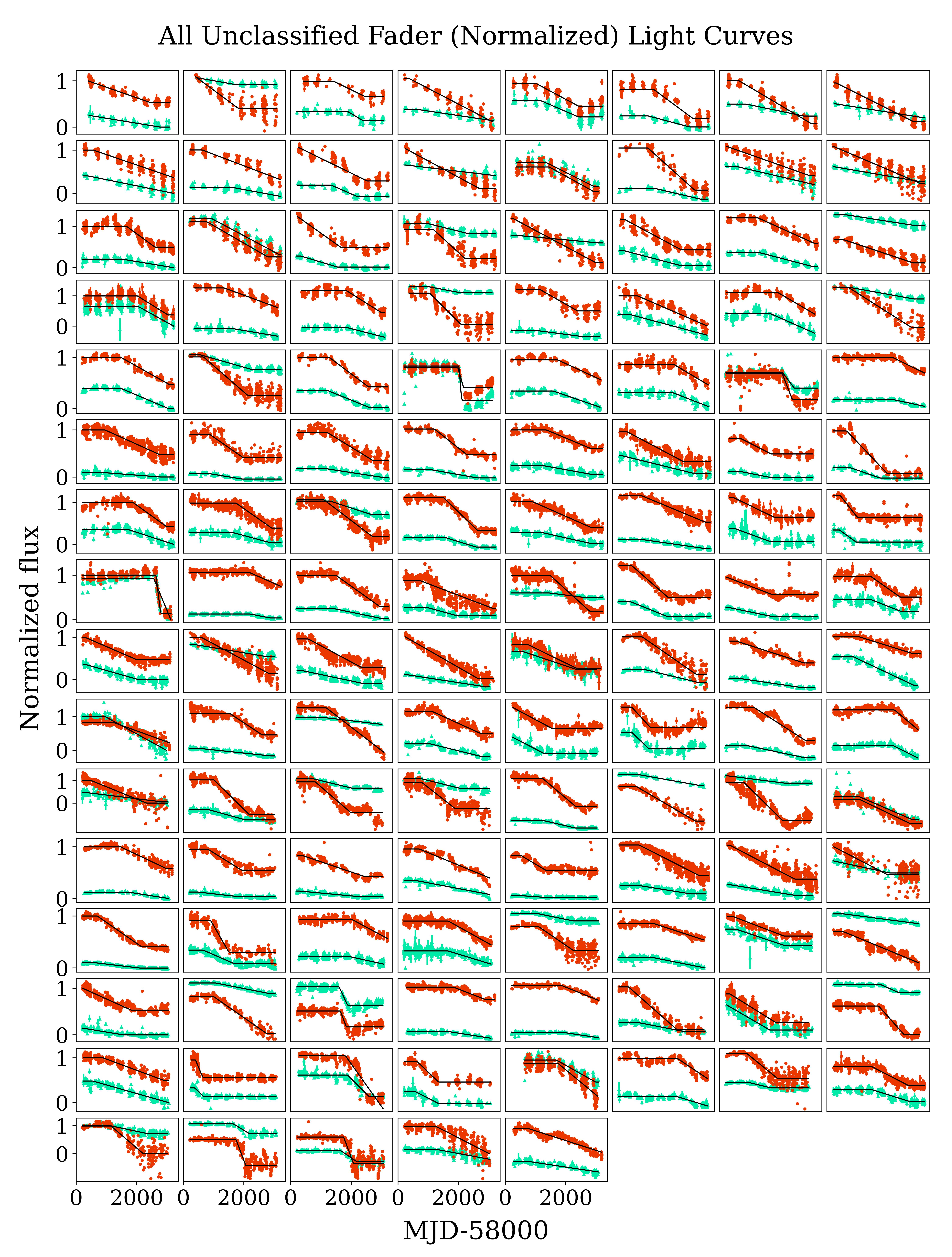}
    \caption{Light curves of all 125 Faders extracted from the search, after all filters. Fluxes are normalized by the maximum and minimum flux of both filters, preserving color information.}
    \label{fig:allcands}
\end{figure*}




The faders found from the ZTF search are listed in Table \ref{tab:candidates}. It includes the coordinates, fractional flux drop, and the timescale of the drop in both bands. Both values are taken directly from the linear model parameters in Equation \ref{eq:16}. Figure \ref{fig:allcands} shows the normalized light curves of all 125 faders. The light curves are normalized by taking the higher $F_0$ of either band as 1 and the lower $F_0-\Delta F$ of either band as 0.

\section{Potential FccSN visibility calculations} \label{app:calcs}

\subsection{Kroupa IMF} \label{app:imf}
The Kroupa IMF used in Sections \ref{ss:formation} and \ref{sec:discussion} is as follows:
\begin{equation}
    \alpha=\begin{cases}
        1.8 & 0.08 \leq M < 0.5\\
        2.7 & 0.5 \leq M < 1.0\\
        2.3 & M \geq 1.0
    \end{cases}
    \label{eq:kroupa alpha}
\end{equation}
\begin{equation}
    \frac{dN}{dM}=\xi_{0, \alpha}M^{-\alpha}
\end{equation}

where $\xi_{0, \alpha}$ is a constant used for the IMF of a specific setting, per $\alpha$ value for continuity \citep{1955ApJ...121..161S,2001MNRAS.322..231K}. The IMF to determine total mass (as used in SFRs) is

\begin{equation}
    M_{\mathrm{tot}}=\int^{200\ \mathrm{M_\odot}}_{0.08\ \mathrm{M_\odot}} M\xi_{0, \alpha}M^{-\alpha}\ dM
\end{equation}

with the integration limits representing the possible zero-age main sequence (ZAMS) masses.

\subsection{Massive stars in ZTF and Milky Way} \label{app:massive star}

Using the IMF model (Appendix \ref{app:calcs}) and density distribution (Section \ref{ss:density}), we first randomly generate all 100 billion stars (a commonly used range for this quantity is $10^{11}{\sim}10^{12}$; \citealt{Franck_Block_Bloh_Bounama_Garrido_Schellnhuber_2001}) within the Milky Way. These generated stars are then put through the visibility filter (Section \ref{ss:vc}) to find number of stars visible in ZTF. The same caveats apply with the star formation simulation in that the density model is not an actual indicator of stellar formation distribution and the dust model is an oversimplification of the asymmetries of dust within the Milky Way. It is still a useful estimation tool; applying the filter to all generated stars will yield $1.5\times10^9$ stars, approximately the actual number of sources in the ZTF search. We use the same method to find the ratio of massive stars ${>}20\ \rm M_\odot$ (that we assume will fail) visible in ZTF to within the Milky Way, from which we obtain 49\%.

\subsection{LSST and Roman predictions} \label{app:predictions}

To calculate potential FccSN rate within a resolution distance, the SFR within that radius is used in conjunction with the Kroupa IMF \citep{1955ApJ...121..161S, 2001MNRAS.322..231K}:

\begin{equation}
    \mathrm{SFR_d}={\int^{200\ \mathrm{M_\odot}}_{0.08\ \mathrm{M_\odot}}\xi_{d, \alpha}M^{-\alpha}\ MdM}
    \label{eq:xi}
\end{equation}

where $\xi_d$ and $\rm SFR_d$ are the $\xi$ constant and SFR at the limiting distance. The number of potential FccSNe, based on SFR, is thus

\begin{equation}
    N_{\mathrm{FccSN}}=\int^{200\ \mathrm{M_\odot}}_{20\ \mathrm{M_\odot}} \xi_{d, \alpha} M^{-\alpha}\ dM.
\end{equation}

For limiting magnitude predictions, the absolute magnitude of a star of mass $M$ (derived through MESA simulations; \citealt{2011ApJS..192....3P}) and the SFR densities of the local universe is combined with the above IMF calculation.

\begin{equation}
    D(M)=10^{\frac{\mathrm{Mag_{lim}}-\mathrm{Mag_{abs}}(M)}{5}+1}
\end{equation}

$D(M)$ is the limiting distance of a star of mass $M$, $\mathrm{Mag_{lim}}$ is the limiting magnitude of the instrument, and $\mathrm{Mag_{abs}}(M)$ is the absolute magnitude of the star $M$. From this distance, one can calculate the predicted number of FccSN at a given instrumental magnitude limit:

\begin{equation}
    \begin{matrix}
        N_{\mathrm{FccSN}}=\sum^{N}_{i=1} [\frac{4}{3}\pi (D(M_i))^3\ \xi_\alpha(D(M_i))\int^{M_{i+1}}_{M_{i}}M^{-\alpha}\ dM]; \\
        M_1=20, M_{N+1}=200.
    \end{matrix}
\end{equation}

Here, $\xi(D(M))$ depends on the limiting distance of the given star mass, and would use the SFR density instead of SFR from Equation \ref{eq:xi}.

\end{document}